\documentclass[showpacs,preprintnumbers,preprint,amsmath,amssymb, graphicx, graphics, epsfig]{revtex4}
\input{epsf}
\usepackage{epsfig}
\usepackage{dcolumn}
\usepackage{amssymb}
\usepackage{bm}
\usepackage{amsmath}

\begin{document}
\def\be{\begin{equation}}
\def\ee{\end{equation}}
\def\bdm{\begin{displaymath}}
\def\edm{\end{displaymath}}
\def\erfc{\hbox{erfc }}
\def\vpa{v_{\parallel }}
\def\vper{v_{\perp }}
\def\Omm{\Omega }
\def\ppa{p_{\parallel }}
\def\pper{p_{\perp }}
\def\ppv{\vec{p}}
\def\kkv{\vec{k}}
\def\omm{\omega}
\def\krz{{\bf \times }}
\def\dele{\delta \vec{E}(\vec{k},\omega )}
\def\delb{\delta \vec{B}(\vec{k},\omega )}
\def\Ab{\sum_a\omm _{p,a}^2(m_ac)^3}
\def\erf{\rm {erf }}

\title{Is the Weibel instability enhanced by the suprathermal populations, or not?}

\author{M. Lazar$^1$  \footnote{E-mail: mlazar@tp4.rub.de}, R. Schlickeiser$^{1,2}$ and
S. Poedts$^3$}
\affiliation{$^1$Research Department - Plasmas with Complex
Interactions, \\Ruhr-Universit\"at Bochum, D-44780 Bochum, Germany \\
$^2$Institut f\"ur Theoretische Physik IV,
Ruhr-Universit\"at Bochum, D-44780 Bochum, Germany \\
$^3$Centre for Plasma Astrophysics, Celestijnenlaan 200B, 3001 Leuven, Belgium}

\date{\today}

\begin{abstract}

The kinetic instabilities of the Weibel-type are presently invoked in a
large variety of astrophysical scenarios because anisotropic
plasma structures are ubiquitous in space. The Weibel instability is
driven by a temperature anisotropy which is commonly modeled by a
bi-axis distribution function, such as a bi-Maxwellian or a
generalized bi-Kappa. Previous studies have been limited to a
bi-Kappa distribution and found a suppression of this instability in
the presence of suprathermal tails. In the present paper it is shown that the Weibel
growth rate is rather more sensitive to the shape of the anisotropic
distribution function. In order to illustrate the distinguishing
properties of this instability a \emph{product-bi-Kappa
distribution} is introduced, with the advantage that this
distribution function enables the use of different values of the spectral
index in the two directions, $\kappa_{\parallel} \ne
\kappa_{\perp}$. The growth rates and the instability threshold are
derived and contrasted with those for a simple bi-Kappa and a
bi-Maxwellian. Thus, while the maximum growth rates reached at the
saturation are found to be higher, the threshold is drastically
reduced making the anisotropic product-bi-Kappa (with small kappas)
highly susceptible to the Weibel instability. This effect could also
rise questions on the temperature or the temperature anisotropy that
seems to be not an exclusive source of free energy for this
instability, and definition of these notions for such Kappa
distributions must probably be reconsidered.

\end{abstract}

\maketitle

\section{Introduction}

The huge amount of incoming radiation and ionized particles  from
space suggests the wide-spread existence of kinetic anisotropies
in cosmic plasmas. Moreover, space
plasmas are sufficiently dilute and their collisionality is sufficiently low (see, e.g., table
8.1 in \cite{s02}) and, therefore, expected to be far from a
Maxwellian equilibrium (only provided by the short-range binary collisions
of plasma particles). Thus, we have direct proofs from the
observations and in-situ measurements that suprathermal populations
are widely present at different altitudes in the solar wind plasma
\cite{fel75, pi87, mak97, fis06} and
probably in the solar corona \cite{pier99}. These populations exhibit suprathermal high
energy tails and are fitted quite well by the family of Kappa distribution
functions \cite{vas68}, which are power laws in particle speed.
The Kappa distribution generalizes the notion of equilibrium in
collisionless plasmas deviated from a thermal (Maxwellian) equilibrium,
and which contain fully developed turbulence in a quasi-stationary
equilibrium \cite{leu02, fis06, treu08}.
In such plasma systems, the temperature is redefined on the basis of a
superadditive entropy \cite{tsal95, treu08}.

The so-called Kappa distributions are defined by using a spectral index
$\kappa$, which determines the slope of the high energy tails in the
velocity spectrum of plasma particles. In the limit of very large
$\kappa \to \infty$, the Kappa functions degenerate into Maxwellians
\cite{st91}. The effects of these suprathermal populations and their
anisotropies on the threshold conditions and the linear growth rates
of kinetic instabilities have been studied by modeling plasma with
diverse anisotropic distributions fully or partially populated by
Kappa particles
\cite{st92,xts93,mace98,xts98,ls00,hm02,tm02,mh03,chm07,zm07,ts08,lss08,lspt08,lp09,basu09,ltsp10}.
These effects largely vary, depending on the shape of the
distribution function and the nature and the frequency of the plasma
mode.  A selection of Kappa distributions and their limiting forms
for a large $\kappa \to \infty$ is given in Ref.~\cite{st91}, Table
I. The most simple anisotropic distribution is the bi-Kappa
distribution function introduced here in Eq.~(\ref{e1}), and which
describes a temperature anisotropy, $T_{\perp} \ne T_{\parallel}$,
with the same index $\kappa_{\parallel} = \kappa_{\perp} = \kappa$
for both directions. This distribution has widely been used for
describing kinetic instabilities in a drifting or a non-drifting
plasma, e.g, the ion acoustic instability \cite{st92}, the
electromagnetic ion-cyclotron and firehose instabilities
\cite{xts93, lp09}, the mirror unstable mode \cite{ls00}, the
whistler and electron cyclotron instabilities \cite{mace98, xts98,
tm02,ts08,lspt08}, and the nonresonant instabilities of Weibel-type
\cite{zm07,lss08,ltsp10}.

Recently, Basu \cite{basu09} has reviewed the stability properties
of hydromagnetic waves in a plasma distributed after a
\emph{product-bi-Kappa} function such as the one introduced here in 
Eq.~(\ref{e4}). The product-bi-Kappa distribution function represents a
generalization of the Kappa-Maxwellian distribution, which is a
product of a one-dimensional Kappa distribution along a preferred
direction in space, e.g., the guiding magnetic field, and a
Maxwellian distribution in the perpendicular plane \cite{hm02}. In a
hybrid Kappa-Maxwellian plasma, unlike a uniform Maxwellian or a
uniform Kappa, the dispersion properties and the stability were
found to be markedly changed \cite{hm02,mh03,chm07}. Notable is that,
because of the anisotropy of the contours in the velocity space,
such a Kappa-Maxwellian distribution can be unstable even for equal
parallel and perpendicular temperatures, or some very well-known
unstable modes become stable to a temperature anisotropy in such
asymmetric distributions \cite{chm07}.

The product-bi-Kappa distribution function has not often been applied
so far, but we think that this function deserves further consideration
for the following reasons: (1) it decouples the dynamics of the plasma
particles over the two principal directions allowing not only for distinct
temperatures, $T_{\parallel}$ and $T_{\perp}$, but for distinct
spectral indices, $\kappa_{\parallel}$ and $\kappa_{\perp}$, as well;
(2) there is a possibility to regulate this
coupling in accord to a guiding magnetic field or other external
constraints by the interplay of $\kappa_{\parallel}$ and
$\kappa_{\perp}$; (3) compared to a bi-Kappa, a product-bi-Kappa
distribution allows for a more realistic description and (4) enlarges
the number of distributions with an elaborated dispersion approach;
(5) one more important feature (recently revealed for a similar
Kappa-Maxwellian plasma \cite{chm07}) is that such a product-bi-Kappa
model permits further analytical progress leading to
tractable expressions for the dielectric tensor elements enabling, for example,
the investigation of oblique modes in a magnetized plasma.

Electromagnetic instabilities of the Weibel-type \cite{w59, f59}
are driven by an arbitrary deviation of the particle velocity distribution
from its equilibrium, whether it is a bulk (relative) motion of streaming
particles \cite{f59} or a temperature anisotropy \cite{w59}.
Kinetic anisotropies are ubiquitous in space plasmas (extending
from heating flows and temperature anisotropies to
particle jets and shock waves in outflows, or interpenetrating plasma
shells in interplanetary wind) and, presently, Weibel instabilities
are invoked in a large variety of astrophysical scenarios.
These instabilities create magnetic field such as the cosmological
seeds necessary for the dynamo mechanism \cite{ss03,lswp09}, or a quasi-stationary magnetic
field boost for the synchrotron emissions in astrophysical sources \cite{ml99}.
Moreover, the magnetic field fluctuations observed in interplanetary space
are also attributed to these instabilities (e.g., filamentation, whistler, mirror,
oblique firehose) \cite{sto06, bale09}, which compete with other constraints
of plasma particles (adiabatic expansion, Coulomb collisions) and
maintain a relatively small temperature anisotropy in the solar wind
\cite{ka02,hel06,st08, bale09, laz10}.

Here we discuss the Weibel instability driven by a temperature anisotropy of
plasma particles. Recent studies of this instability have been limited to a
bi-Kappa distribution and found that the instability
is suppressed in the presence of the Kappa tails \cite{zm07, lss08}.
For the same arguments formulated above, we investigate here the product-bi-Kappa
distribution as a favorable alternative that seems to add an excess of
free energy in the velocity space, and to enhance the instability.
In the present paper we proceed to a comparative analysis of the effects of
these two distributions, the bi-Kappa and the product-bi-Kappa, on the Weibel
instability. The plasma is assumed to be collisionless  and spatially homogeneous.

In order to describe the initially unperturbed plasma system we
first introduce the bi-Kappa distribution function
\begin{align}
F_1 (v_{\parallel}, v_{\perp}) = {1 \over \pi^{3/2} \theta_{\perp}^2
\theta_{\parallel}} \, {\Gamma[\kappa +1] \over \kappa^{3/2}
\Gamma[\kappa -1/2]} \left(1 + {v_{\parallel}^2\over \kappa
\theta_{\parallel}^2 } + {v_{\perp}^2\over \kappa \theta_{\perp}^2
}\right)^{-\kappa-1}, \label{e1}
\end{align}
using polar coordinates $(v_x, v_y, v_z) = (v_{\perp} \cos \phi,
v_{\perp} \sin \phi, v_{\parallel})$ in the particle velocity space.
This distribution is normalized to unity, $\int d^3v\, F_1 = 1$,
and makes reference to the equivalent thermal velocities
$\theta_{\parallel, \perp}$, which relate to the effective
temperatures of the plasma particles:
\begin{align}
T_{\parallel} & \equiv {m v_{T_\parallel}^2 \over 2 k_B} = {m \over k_B} \int d{\bf v} v_{\parallel}^2
F_1 (v_{\parallel}, v_{\perp}) =  {m \over 2 k_B} \, {2 \kappa
\over 2 \kappa -3} \theta_{\parallel}^2 \label{e2}  \\
 T_{\perp} & \equiv {m v_{T_\perp}^2 \over 2 k_B}  = {m \over 2 k_B} \int d{\bf v} v_{\perp}^2
F_1 (v_{\parallel}, v_{\perp}) =  {m \over 2 k_B} \, {2 \kappa
\over 2 \kappa -3} \theta_{\perp}^2 \label{e3},
\end{align}
for a spectral index $\kappa > 3/2$.

\begin{figure} \centering
   \includegraphics[width=140mm]{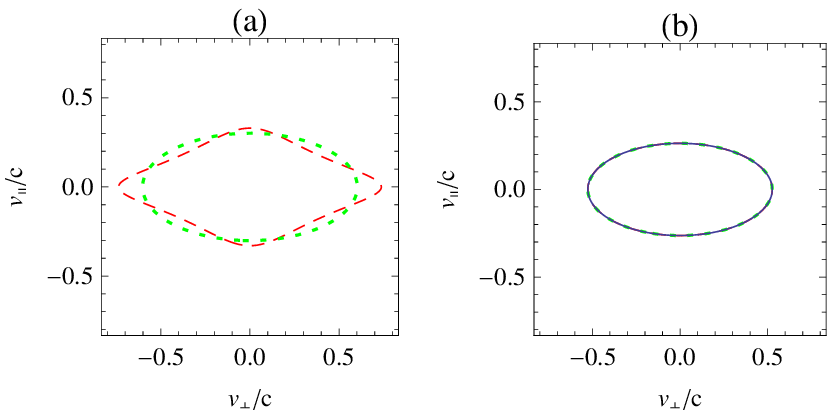} \\
   \includegraphics[width=140mm]{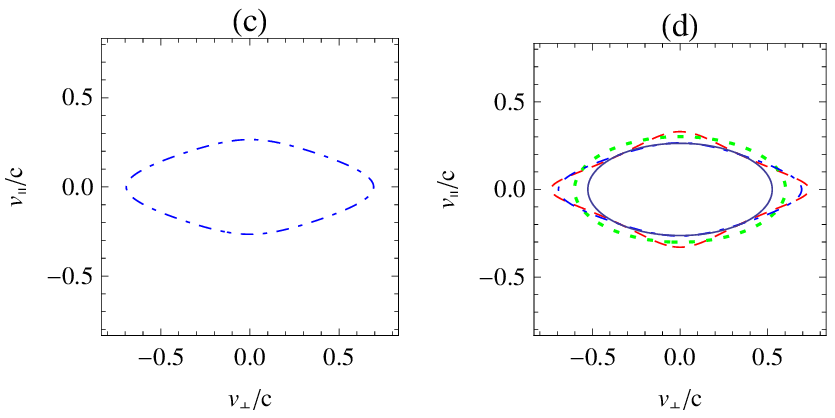}
    \caption{Contours of the distribution functions (\ref{e1})
    and (\ref{e4}) are plotted with dotted and dashed lines, respectively,
    in panel (a) for $\kappa_{\parallel} = \kappa_{\perp} = \kappa = 3$ and
$v_{T_{\perp}}/c =2 v_{T_{\parallel}}/c = 0.2$, and in panel (b) for
a very large $\kappa \to \infty$ leading to an exact fit of these two contours
with that for a bi-Maxwellian (solid line). A Kappa-Maxwellian distribution
function (\ref{e4}) with $\kappa_{\parallel} = 3$ and
$\kappa_{\perp} \to \infty$ (dot-dashed line) is plotted in (c), and
all these distributions are compared in panel (d). For a small
Kappa, remark the high energy tails for both distributions and the
prominent anisotropy of the distribution (\ref{e4}).} \label{fig1}
\end{figure}

The second distribution function examined here is a \emph{product-bi-Kappa}
\begin{align}
F_2 (v_{\parallel}, v_{\perp}) = {1 \over \pi^{3/2} \theta_{\perp}^2
\theta_{\parallel}} \, {\Gamma[\kappa_{\parallel} +1] \over
\kappa_{\parallel}^{1/2} \Gamma[\kappa_{\parallel} +1/2]} \left(1 +
{v_{\parallel}^2\over \kappa_{\parallel} \theta_{\parallel}^2
}\right)^{-\kappa_{\parallel} -1} \left(1 + {v_{\perp}^2\over
\kappa_{\perp} \theta_{\perp}^2 }\right)^{-\kappa_{\perp}-1}
\label{e4}
\end{align}
as was defined in Summers and Thorne \cite{st91}. This function is
also normalized to unity $\int d^3v\, F_2 = 1$, and the equivalent
thermal velocities $\theta_{\parallel, \perp}$ are given by
\begin{align}
T_{\parallel} & \equiv {m v_{T_\parallel}^2 \over 2 k_B} = {m \over k_B} \int d{\bf v} v_{\parallel}^2
F_1 (v_{\parallel}, v_{\perp}) =  {m \over 2 k_B} \, {2 \kappa_{\parallel}
\over 2 \kappa_{\parallel} -1} \theta_{\parallel}^2 \label{e5} \\
 T_{\perp} & \equiv {m v_{T_\perp}^2 \over 2 k_B}  = {m \over 2 k_B} \int d{\bf v} v_{\perp}^2
F_1 (v_{\parallel}, v_{\perp}) =  {m \over 2 k_B} \, {\kappa_{\perp}
\over \kappa_{\perp} -1} \theta_{\perp}^2 \label{e6},
\end{align}
for different spectral indices $\kappa_{\perp} > 1$ and
$\kappa_{\parallel}> 1/2$, respectively.

Contour plots of the distribution functions (\ref{e1})
and (\ref{e4}) are displayed in Fig.~\ref{fig1}.
At low values of $\kappa$, there are significant differences between these two distributions
and a prominent asymmetry of the product-bi-Kappa distribution, which
indicate a surplus of temperature anisotropy, e.g., in panels (a) and (d).
However, both these distributions functions approach the same bi-Maxwellian in the limit of a very
large spectral index, e.g., in panel (b).

Assuming an excess of perpendicular temperature ($T_{\perp} >
T_{\parallel}$), the Weibel instability develops along
the parallel direction (subscript "$\parallel$") with a
wave-number $k = k_{\parallel}$, aperiodic, Re($\omega$)$
\equiv \omega_r = 0$, and with a growth rate
Im($\omega$)$ \equiv \omega_i > 0$ . By using
standard techniques based on the linearized Vlasov-Maxwell equations,
it is easy to find the dispersion relation for the
electromagnetic modes propagating along the parallel direction \cite{zm07,lss08}
\be {\omega^2 - k^2c^2 \over \omega_p^2} -1 +\pi
k\int_{-\infty}^{\infty} { dv_{\parallel} \over \omega -k
v_{\parallel}} \int_0^{\infty} dv_{\perp} v_{\perp}^3 {\partial F
\over \partial v_{\parallel}} = 0, \label{e7}, \ee
where $\omega_p = (4 \pi n_e e^2 /m_e)^{1/2}$ is the plasma
frequency.

\begin{figure}[h] \centering
   \includegraphics[width=60mm]{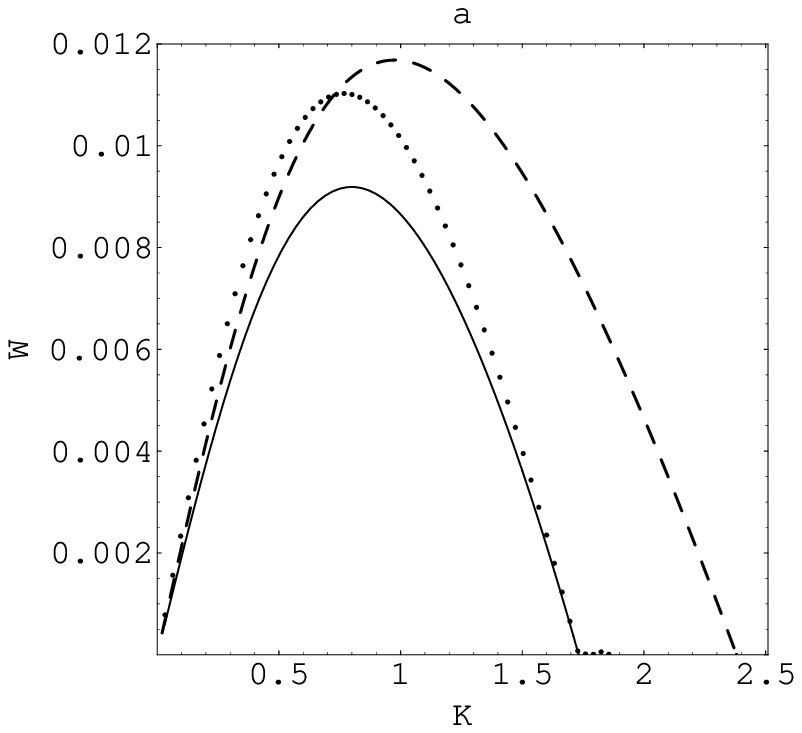} $\;\;\;$
   \includegraphics[width=60mm]{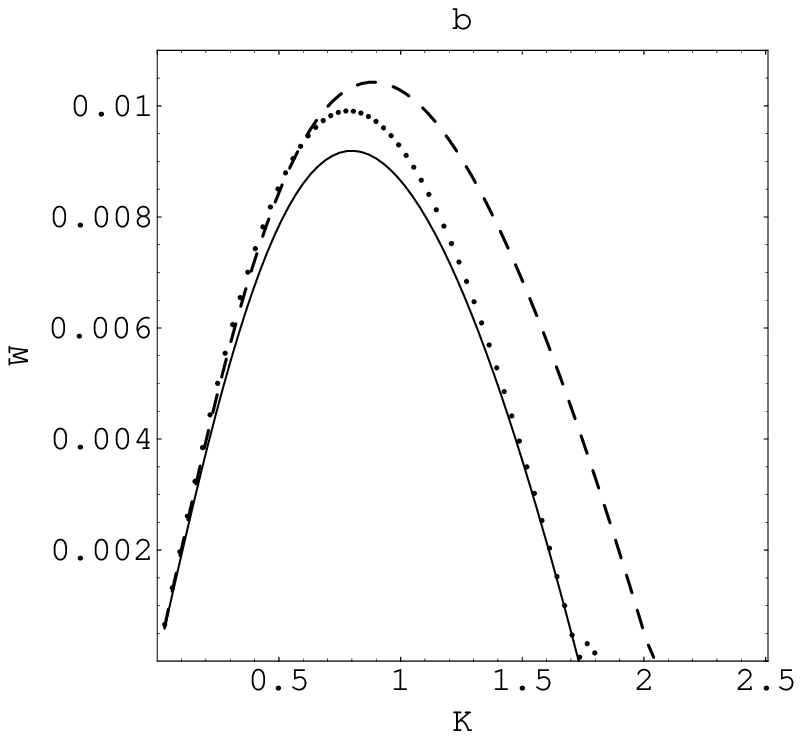} $\;\;\;$
   \includegraphics[width=60mm]{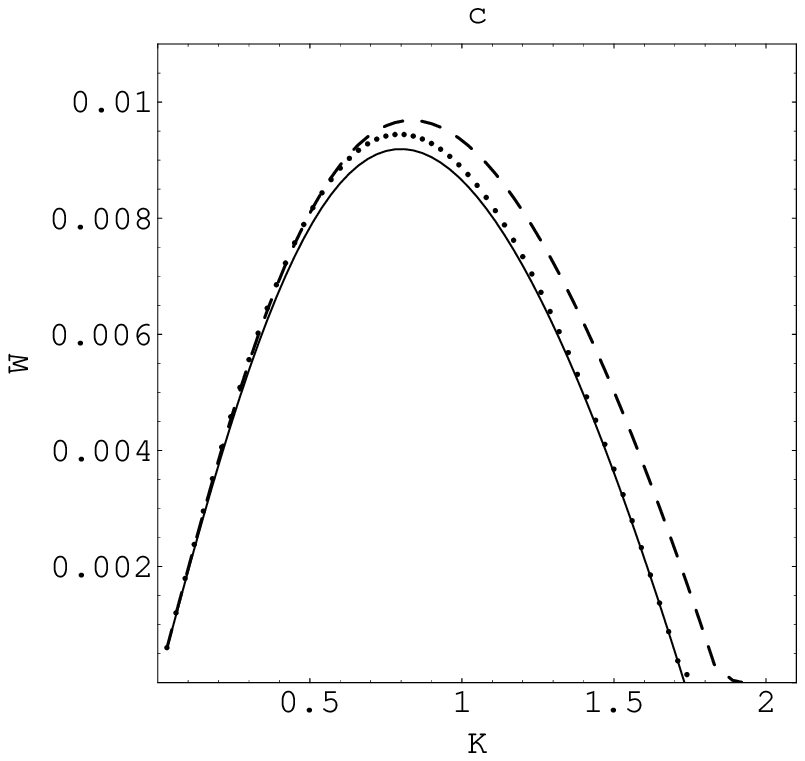}
    \caption{The growth rates of the Weibel instability, solutions of Eqs. (\ref{e8})
    (solid lines) and (\ref{e11}) (dashed lines) for an anisotropic Kappa
    distributed plasma with $ v_{T_{\perp}}/v_{T_{\parallel}} = 2$, $v_{T_{\parallel}} = 0.02 c$
    and for three values of $\kappa_{\parallel} = \kappa = 2$ (in a), $\kappa = $4 (in b), and
    $\kappa = $ 10 (in c).
   The growth rates for a bi-Maxwellian plasma ($\kappa \to \infty$) are plotted with dotted lines.} \label{fig2}
\end{figure}

First we insert the bi-Kappa distribution function (\ref{e1}) in
Eq.~(\ref{e7}) and find a dispersion relation \cite{zm07,lss08}
\begin{align} 0& = {\omega^2 - k^2c^2 \over \omega_p^2} -1 + {\theta_{\perp}^2
\over \theta_{\parallel}^2 }\, \left[1 +  {\omega \over k
\theta_{\parallel}} \, Z_{\kappa} \left({\omega \over k
\theta_{\parallel}} \right) \right] \notag \\
&= {\omega^2 - k^2c^2 \over \omega_p^2} -1 + {T_{\perp} \over
T_{\parallel}}\, \left[1 +  {\omega \over k \theta_{\parallel}} \,
Z_{\kappa} \left({\omega \over k \theta_{\parallel}} \right) \right]
, \label{e8}
\end{align}
in terms of the modified plasma dispersion function
\be Z_{\kappa}(f) = {1 \over \pi^{1/2} \kappa^{1/2}} \, {\Gamma
[\kappa] \over \Gamma \left[\kappa - 1 / 2\right]} \,
\int_{-\infty}^{+\infty} dx \, {(1+x^2/\kappa)^{- \kappa } \over x -
f}, \; \;\; \Im (f) > 0. \label{e9} \ee
This dispersion relation admits purely growing solutions of the Weibel type for wave-numbers
smaller than a cutoff value
\be k_{c1} = {\omega_p \over c}\, \left({\theta_{\perp}^2 \over
\theta_{\parallel}^2} -1 \right)^{1/2} = {\omega_p \over c}\,\left(
{T_{\perp} \over T_{\parallel}} -1\right)^{1/2}.\label{e10} \ee
The exact numerical growth rates are displayed in Fig.~\ref{fig2}
with solid lines for small values of $\kappa$, and with dotted lines
for a very large $\kappa \to \infty$ (bi-Maxwellian plasma). In this
case the cutoff wave-number does not depend on the spectral index
$\kappa > 3/2$ and this is confirmed in Fig.~\ref{fig2} where
$k_{c1} c /\omega_p=  \left(T_{\perp}/T_{\parallel} - 1\right)^{1/2}
\simeq 1.73$ is the same for all three cases.  For the existence
of a finite $k_{c1} \ne 0$, the instability threshold $\tau = \min
(T_{\perp}/T_{\parallel})$ is simply found as $T_{\perp} /
T_{\parallel} > \tau = 1$.

If we insert the second distribution function (\ref{e4}), the
dispersion relation (\ref{e7}) takes a new and different form
\begin{align} 0& = {\omega^2 - k^2c^2 \over \omega_p^2} -1 + {\theta_{\perp}^2
\over \theta_{\parallel}^2 }\, {\kappa_{\perp}(\kappa_{\parallel}
+1/2) \over \kappa_{\parallel}(\kappa_{\perp} -1)}\,\left[1 +
{\omega \over k \theta_{\parallel}} \, Z_{\kappa_{\parallel}}
\left({\omega \over k \theta_{\parallel}} \right) \right] \notag \\
& = {\omega^2 - k^2c^2 \over \omega_p^2} -1 + {T_{\perp} \over
T_{\parallel}}\, {\kappa_{\parallel} +1/2 \over \kappa_{\perp} -1/2}
\, \left[1 + {\omega \over k \theta_{\parallel}} \,
Z_{\kappa_{\parallel}} \left({\omega \over k \theta_{\parallel}}
\right) \right], \label{e11} \end{align}
where
\be Z_{\kappa_{\parallel}}(f) = {1 \over \pi^{1/2}
\kappa_{\parallel}^{1/2}} \, {\Gamma [\kappa_{\parallel}+2] \over
\Gamma \left[\kappa_{\parallel} +3 / 2\right]} \,
\int_{-\infty}^{+\infty} dx \, {(1+x^2/\kappa_{\parallel})^{-
\kappa_{\parallel} -2 } \over x - f}, \; \;\; \Im (f) > 0.
\label{e12} \ee
This dispersion relation does not involve any dependence on $\kappa_{\perp}$
but only on $\kappa_{\parallel}$, and it can therefore be further applied
to describe the instability of a Kappa-Maxwellian (one-dimensional Kappa
distribution in the parallel direction and a Maxwellian distribution in the
perpendicular plane). Notice that, both the plasma dispersion functions
(\ref{e9}) and (\ref{e12}) approach the standard dispersion function of
Fried and Conte \cite{fc61} in the limit of a very large
$\kappa, \kappa_{\parallel} \to \infty$.

The aperiodic solutions of dispersion relation (\ref{e11}) are displayed
in Fig.~\ref{fig2} with dashed lines.
In this case not only the growth rates increase (as was expected from
Fig.~\ref{fig1}, where the excess of anisotropy of a product-bi-Kappa
distribution compared to a bi-Kappa, is evident), but the instability
extends as well to larger wave-numbers (smaller wave-lengths) due to
a non-negligible dependence on the spectral index $\kappa_{\parallel}$
of the new cutoff wave-number
\be k_{c2} = {\omega_p \over c}\, \left[{T_{\perp} \over
T_{\parallel}}\, {\kappa_{\parallel}+1/2 \over
\kappa_{\parallel}-1/2} -1\right]^{1/2}.\label{e13} \ee
If we look, for example, at the growth rates displayed with dashed line
in Fig.~\ref{fig2}a, and which are solutions of Eq.~(\ref{e11})
for an index $\kappa_{\parallel} = \kappa = 2$, and a
temperature anisotropy $v_{T_{\perp}}/v_{T_{\parallel}} = 2$ (where
$v_{T_{\perp, \parallel}} = \sqrt{2 k_B T_{\perp, \parallel}/m}$),
the cutoff wave-number is indeed larger $k_{c2} c /\omega_p=
\left[5v_{T_{\perp}}^2/(3 v_{T_{\parallel}}^2) - 1\right]^{1/2}
\simeq 2.38 > k_{c1} \simeq 1.73$. The cutoff wave-number
(\ref{e13}) decreases and approaches (\ref{e10}) only for a
sufficiently large $\kappa \to \infty$.

\begin{figure}[h] \centering
  \includegraphics[width=90mm]{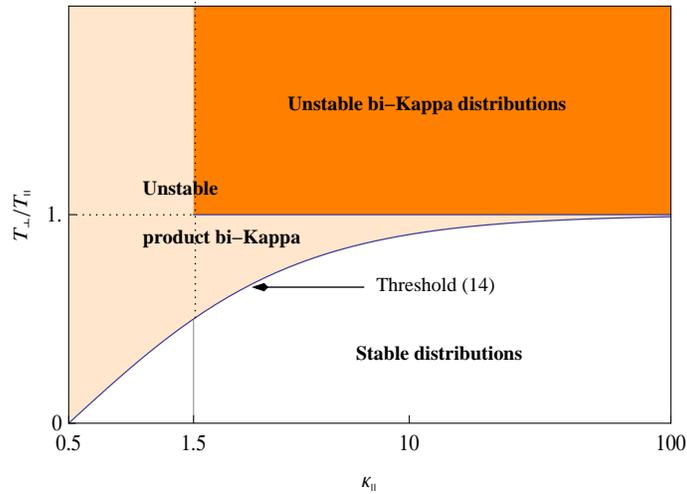}
   \caption{Contrast of the instability thresholds showing the unstable bi-Kappa
distribution in the gray (orange online) region in the limits of
$\kappa_{\parallel} > 1.5$ and
   $T_{\perp} / T_{\parallel} > 1$, and the unstable product-bi-Kappa distribution
in the light-gray (light-orange online) region limited by
$\kappa_{\parallel} > 0.5$ and the condition (\ref{e14}) for the
temperature anisotropy, and including the more restrained gray
region.} \label{fig3}
\end{figure}

Furthermore, for a product-bi-Kappa, the anisotropy threshold $\tau$
decreases to less than unity
\be
{T_{\perp} \over T_{\parallel}} > \tau
= {\kappa_{\parallel} -1/2 \over \kappa_{\parallel} + 1/2} < 1. \label{e14}
\ee
and towards the limit of $\kappa_{\parallel} \to 1/2$, the
threshold vanishes, $\tau \sim 0 $, and the instability can grow
freely. If $\kappa_{\parallel} = 2$ the instability will develop for
$T_{\perp}/T_{\parallel} > \tau= 3/5$, and then the instability
threshold increases asymptotically to unity, $\tau \to 1$, as the
spectral index becomes very large $\kappa_{\parallel} \to \infty$.

To conclude, in this paper we have revised the effects of Kappa
anisotropic distributed plasmas on the Weibel instability. Previous
models were limited to a bi-Kappa distribution and have been extended
by introducing a more general \emph{product-bi-Kappa} distribution. In this
new distribution, the dynamics of plasma particles over the two
principal directions are decoupled and characterized by two distinct
temperatures and two distinct spectral indices, $\kappa_{\perp} \ne
\kappa_{\parallel}$. The growth rates and the instability threshold
have been found to be very sensitive to the shape of the anisotropic
distribution. While for a bi-Kappa distribution the Weibel
instability is suppressed, for a product-bi-Kappa the growth rates
are enhanced and the instability threshold is significantly lowered by
comparison to a Maxwellian. 

The enhancing effect can not be attributed to a resonant
interaction with the energetic particles from Kappa tails because
this instability is nonresonant, but it is fully supported by an
excess of anisotropy and free energy of the product-bi-Kappa
distribution in the velocity space. In this sense, the contour plots
from Fig. \ref{fig1} are very suggestive: if we compare contours in 
panels (a) and (b), the bi-Kappa is less asymmetric than the 
bi-Maxwellian leading to lower growth rates of the Weibel instability,
while the product-bi-Kappa is more asymmetric than the bi-Maxwellian 
and enhances the instability.

The temperature anisotropy threshold for the Weibel instability can
be drastically reduced in anisotropic plasmas with
product-bi-Kappa distribution functions that suggests two
possible explanations: either the fundamental notions, such as the
temperature or the temperature anisotropy must be redefined for
these Kappa distributions, or the temperature anisotropy is not an
exclusive source of free energy for this instability.

These results can easily be extended to high beta plasmas ($\beta$ =
thermal energy/magnetic energy) widely present in space and where
the ambient magnetic field has only a minor influence without changing the
essential features of the instability, but sustaining such
asymmetric distributions with $\kappa_{\perp} > \kappa_{\parallel}$
due to some equilibration and isotropization in the perpendicular
plane, and a preferential motion and acceleration along the magnetic
field.

\acknowledgements {This work was supported in part by the Research Department
- Plasmas with Complex Interactions, Ruhr-Universit\"at Bochum,
the Katholieke Universiteit Leuven through the fellowship F/07/061, and 
by the Deutsche Forschungsgemeinschaft (DFG), grant Schl 201/21-1.}


\begin{references}

\bibitem{s02} R. Schlickeiser, {\it Cosmic Ray Astrophysics} (Springer, Heidelberg,
2002).

\bibitem{fel75} W.\,C. Feldman, J.\,R. Asbridge, S.\,J. Bame, M.\,D.
Montgomery, and S.\,P. Gary, J. Geophys. Res. {\bf 80}, 4181 (1975).
\bibitem{pi87} W.\,G. Pilipp, H. Miggenrieder, M.\,D. Montgomery,
K.\,H. Muhlhauser, H. Rosenbauer, and R Schwenn, J. Geophys. Res. {\bf 92}, 1075 (1987).
\bibitem{mak97} M. Maksimovic, V. Pierrard, and P. Riley, Geophys. Res. Let. {\bf 24}, 1151 (1997).
\bibitem{fis06} L.\,A. Fisk and G. Gloeckler, Astrophys. J. {\bf 640}, L79 (2006).
\bibitem{pier99} V. Pierrard, M. Maksimovic, and J.\,F. Lemaire,
J. Geophys. Res. {\bf 104}, 17021 (1999).

\bibitem{vas68} V.\,M. Vasyliunas, J. Geophys. Res. {\bf 73}, 2839 (1968).
\bibitem{leu02} M.\,P. Leubner, Astrophys. Space Sci. {\bf 282}, 573 (2002).
\bibitem{treu08} R.\,A. Treumann and C.\,H. Jaroschek, Phys. Rev. Lett.
{\bf 100}, 155005 (2008).
\bibitem{tsal95} C. Tsallis, Phys. A {\bf 221}, 277 (1995).

\bibitem{st91} D. Summers and R.\,M. Thorne, Phys. Fluids {\bf B 3}, 1835 (1991).
\bibitem{st92} D. Summers and R.\,M. Thorne, J. Geophys. Res. {\bf 97} 16827 (1992).
\bibitem{xts93} S. Xue, R.\,M. Thorne, and D. Summers, J. Geophys. Res. {\bf 98},
17475 (1993); {\bf 101}, 15457 (1996); {\bf 101}, 15467 (1996).
\bibitem{mace98} R.\,L. Mace, J. Geophys. Res. {\bf 103}, 14643 (1998).
\bibitem{xts98} F. Xiao, R.\,M. Thorne, and D. Summers, Phys. Plasmas {\bf 5}, 2489 (1998).
\bibitem{ls00} M.\,P. Leubner and N. Schupfer, J. Geophys. Res. {\bf 105}, 27387 (2000).
\bibitem{hm02} M.\,A. Hellberg and R.\,L. Mace, Phys. Plasmas {\bf 9}, 1495 (2002).
\bibitem{tm02} A.\,K. Tripathi and K.\,D. Misra, Earth Moon Planet. {\bf 88}, 131 (2002)
\bibitem{mh03} R.\,L. Mace and M.\,A. Hellberg, Phys. Plasmas {\bf 10}, 21 (2003).
\bibitem{chm07} T. Cattaert, M.\,A. Hellberg, and R.\,L. Mace, Phys. Plasmas
{\bf 14}, 082111 (2007).
\bibitem{zm07} S. Zaheer and G. Murtaza, Phys. Plasmas {\bf 14}, 022108 (2007).
\bibitem{ts08} A.\,K. Tripathi and R.\,P. Singhal, Planet. Space Sci. {\bf 56}, 310 (2008).
\bibitem{lss08} M. Lazar, R. Schlickeiser, and P.\,K. Shukla, Phys. Plasmas {\bf 15}, 042103 (2008).
\bibitem{lspt08} M. Lazar, R. Schlickeiser, S. Poedts, and R. Tautz, Mon. Not.
R. Astron. Soc. {\bf 390}, 168 (2008).
\bibitem{lp09} M. Lazar and S. Poedts, Astron. Astrophys. {\bf 494}, 311 (2009).
\bibitem{basu09} B. Basu, Phys. Plasmas {\bf 16}, 052106 (2009).
\bibitem{ltsp10} M. Lazar, R. Tautz, R. Schlickeiser, and S. Poedts, Mon. Not.
R. Astron. Soc. {\bf 401}, 362 (2010).

\bibitem{w59}  E.\,S. Weibel, Phys. Rev. Lett. {\bf 2}, 83 (1959).
\bibitem{f59} B.\,D. Fried, Phys. Fluids {\bf 2}, 337 (1959).
\bibitem{ss03} R. Schlickeiser and P.K. Shukla, Astrophys. J. {\bf 599}, L57 (2003).
\bibitem{lswp09} M. Lazar, R. Schlickeiser, R. Wielebinski, and 
S. Poedts, Astrophys. J. {\bf 693}, 1133 (2009).
\bibitem{ml99} M. Medvedev and A. Loeb, Astrophys. J. {\bf 526}, 697 (1999).
\bibitem{sto06} A. Stockem, I. Lerche, and R. Schlickeiser, Astrophys.
J. {\bf 651}, 584 (2006).
\bibitem{bale09} S.\,D. Bale, J.\,C. Kasper, G.\,G. Howes, E. Quataert,
C. Salem, and D. Sundkvist, Phys. Rev. Lett. {\bf 103}, 211101 (2009).
\bibitem{ka02} J.\,C. Kasper, A.\,J. Lazarus, and S.\,P. Gary, Geophys. Res.
Lett. {\bf 29}, 1839 (2002).
\bibitem{hel06} P. Hellinger, P. Travnicek, J.\,C. Kasper, and A.\,J. Lazarus,
Geophys. Res. Lett. {\bf 33}, L09101 (2006).
\bibitem{st08} S. Stverak, P. Travnicek, M. Maksimovic, et al., J.
Geophys. Res. {\bf 113}, A03103 (2008).
\bibitem{laz10} M. Lazar, S. Poedts, and R. Schlickeiser, AIP Proceedings (2010) in press.

\bibitem{fc61} B.\,D. Fried and S.\,D. Conte, {\it The Plasma Dispersion Function} 
(Academic Press, New York, 1961).

\end{references}
\end{document}